# The Infrared Imaging Spectrograph (IRIS) for TMT: Multi-tiered Wavefront Measurements and Novel Mechanical Design

Jennifer Dunn*[a], David Andersen[a], Edward Chapin[a], Vlad Reshetov[a],
Ramunas Wierzbicki[a], Glen Herriot[a], Dean Chalmers[b], Victor Isbrucker[c],
James E. Larkin[d], Anna M. Moore[e], Ryuji Suzuki[f]

[a]National Research Council Herzberg, 5071 W. Saanich Rd., Victoria, BC, V9E 2E7, Canada.
[b]National Research Council Herzberg, 717 White Lake Rd, Cawston, BC V0X 1C0, Canada.
[c]Isbrucker Consulting, 4 Third Street, Sturgeon Point, Ontario, K0M 1N0, Canada.
[d]Department of Physics and Astronomy, University of California, Los Angeles, CA 90095-1547.
[e]Caltech Optical Observatories,1200 E California Blvd., Mail Code 11-17, Pasadena, CA 91125.
[f]Advanced Technology Center, National Astronomical Observatory of Japan,
2-21-1 Osawa, Mitaka, Tokyo, 181-8588, Japan.

## ABSTRACT

The InfraRed Imaging Spectrograph (IRIS) will be the first light adaptive optics instrument on the Thirty Meter Telescope (TMT). IRIS is being built by a collaboration between Caltech, the University of California, NAOJ and NRC Herzberg. In this paper we present novel aspects of the Support Structure, Rotator and On-Instrument Wavefront Sensor systems being developed at NRC Herzberg. IRIS is suspended from the bottom port of the Narrow Field Infrared Adaptive Optics System (NFIRAOS), and provides its own image de-rotation to compensate for sidereal rotation of the focal plane. This arrangement is a challenge because NFIRAOS is designed to host two other science instruments, which imposes strict mass requirements on IRIS. As the mechanical design of all elements has progressed, we have been tasked with keeping the instrument mass under seven tonnes. This requirement has resulted in a mass reduction of 30 percent for the support structure and rotator compared to the most recent IRIS designs. To accomplish this goal, while still being able to withstand earthquakes, we developed a new design with composite materials. As IRIS is a client instrument of NFIRAOS, it benefits from NFIRAOS's superior AO correction. IRIS plays an important role in providing this correction by sensing low-order aberrations with three On-Instrument Wavefront Sensors (OIWFS). The OIWFS consists of three independently positioned natural guide star wavefront sensor probe arms that patrol a 2-arcminute field of view. We expect tip-tilt measurements from faint stars within the IRIS imager focal plane will further stabilize the delivered image quality. We describe how the use of On-Detector Guide Windows (ODGWs) in the IRIS imaging detector can be incorporated into the AO correction. In this paper, we present our strategies for acquiring and tracking sources with this complex AO system, and for mitigating and measuring the various potential sources of image blur and misalignment due to properties of the mechanical structure and interfaces.

## 1. INTRODUCTION

The Infrared Imaging Spectrograph (IRIS) will be the first light adaptive optics (AO) instrument for the Thirty Meter Telescope (TMT). IRIS is being built in a collaboration between Caltech, the University of California, NAOJ and National Research Council - Herzberg (NRC-H). Currently in its Preliminary Design Phase, IRIS consists of a cryostat, measuring approximately 4.5 m in length, which houses a wide field imager and an integral field spectrograph (IFS), both covering near-infrared (NIR) wavelengths from 0.84 µm to 2.4 µm. Above the cryostat is an On-Instrument Wavefront Sensor (OIWFS), which consists of three fixed detectors fed from configurable probe arms that patrol a ~2-arcmin diameter field-of-view (FOV). IRIS is mounted below and receives light from the facility TMT AO system, the Narrow Field Infrared Adaptive Optics System (NFIRAOS). The OIWFS are an important element of the AO system and are used to sense tip/tilt (TT), focus and plate scale modes. Each OIWFS can be independently configured in a tip/tilt (TT, up to two probes) or tip/tilt/focus (TTF, one probe) mode. These OIWFS are designed to work at the diffraction limit and operate in the NIR (over a 1-2.3 micron broad band) to benefit from the image sharpening provided by NFIRAOS. This OIWFS architecture was chosen to be able to take advantage of relatively faint stars (J<22) and therefore achieve very high sky coverage; TMT requires that NFIRAOS+IRIS achieve excellent image quality for 50% of fields located near the Galactic Pole. In order to compensate for sidereal rotation, the mechanical interface to NFIRAOS incorporates a rotator. NRC-H is responsible for the design of the instrument rotator, handling cart, the OIWFS probe arms, and both the support and interface to NFIRAOS. For an overview of the instrument, see Ref. [1].

---

* Jennifer.Dunn@nrc-cnrc.gc.ca phone 1 250 363-6912; fax 1 250 363-0045; www.hia-iha.nrc-cnrc.gc.ca

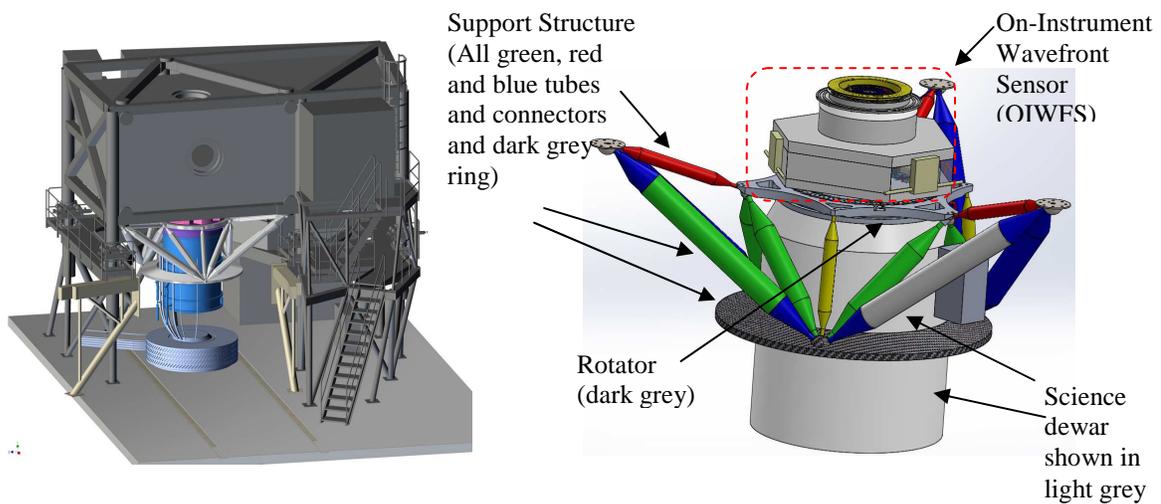

**Figure 1 –** *left*: **IRIS in blue and pink, mounted to the NFIRAOS bottom client port (the side and top ports can also be seen as circular indentations).** *right*: **IRIS instrument showing the support structure, on-instrument wavefront sensor (OIWFS), rotator and science dewar.**

This paper focuses on updates to the support structure that connects IRIS to NFIRAOS since the Conceptual Design Phase (CDP). We also summarize our method for synchronizing the low-order wavefront measurements provided to NFIRAOS from the OIWFS, and up to four On-Detector Guide Windows (ODGW, in which TT information is obtained directly from the Imager). Coordinating this feedback is a challenge, as the detectors in question are triggered by NFIRAOS (and must therefore be interleaved with the independently-controlled science exposures in the case of ODGWs), and must provide data in a timely manner to the NFIRAOS Real Time Controller (RTC) which operates the real-time AO control loops.

## 2. SUPPORT STRUCTURE/ROTATOR

During the Conceptual Design Phase (CDP) in 2011, the structure and rotator used to connect IRIS to NFIRAOS was located at the top of the instrument using mounting pads. However, structural design changes within NFIRAOS affected the location to which IRIS will be attached, and as a result, the rotator was moved from the top of IRIS to its midline, as shown Figure 2. At this new location, the size and weight of the rotating bearing increased significantly. Furthermore, earthquake survivability came into question, as the fundamental frequency of this larger mass was found to be unacceptably low.

During the Preliminary Design Phase (PDP) we have undertaken two major design challenges to address these issues:
   (i)   to decrease the mass over our initial design by approximately 32% in order to fit within a strict mass allocation of 6800 kg (noting that NFIRAOS, IRIS, and two other NFIRAOS client instruments will ultimately be suspended by the telescope structure);
   (ii)  and to increase the rigidity of the mounting structure, thereby further raising the fundamental frequency of the instrument. For the Instrument rotator and support structure the first mode must be larger than 25Hz.

|  | December 2014 Weight (kg) | Current Weight (kg) |
|---|---|---|
| Instrument rotator | 2500 | 1900 |
| Support Structure | 1200 | 800 |
| Total | 3700 | 2700 |
| Current Mass Goal | 2500 | |

To achieve these goals, a wide variety of materials and structures were considered, including: steel, aluminum, a carbon fiber frame and shell, and high modulus carbon fiber (see Figure 3). Throughout this effort, the function of the supporting structure was decoupled from that of the rotator. The following subsection details our solution.

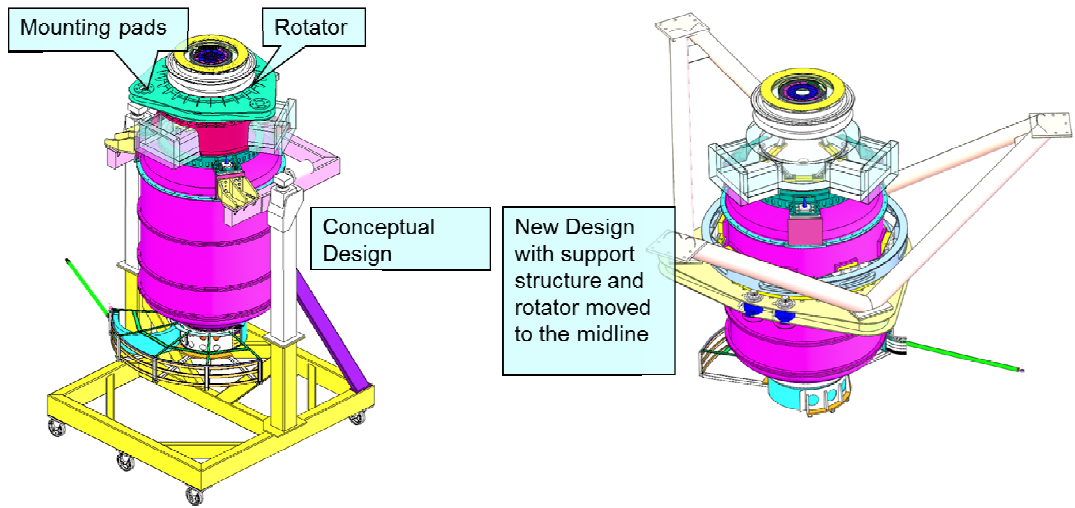

**Figure 2 –** *left:* **Pre-Conceptual design of IRIS with the rotator at the top, in which the only interface structure required was the mounting pads.** *right:* **Design at Conceptual design with support arms and rotator around the mid-section of IRIS.**

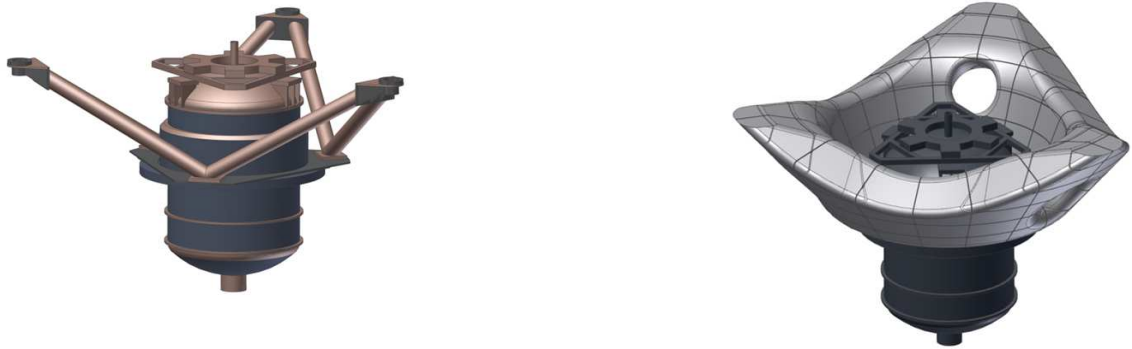

**Figure 3 - Different support structure concepts considered during the drive to lose weight.** *left:* **a carbon fiber replacement using struts.** *right:* **a solid molded carbon shell structure mount.**

### 2.1 DETAILS OF SUPPORT STRUCTURE/ROTATOR

At the time of the conceptual design the mounting interface to NFIROAS (V1 in Figure 4) consisted of three mounting pads equally spaced on a 1500 mm inscribed diameter circle. The interface structure and rotator concept then consisted of a monolithic triangular frame with mounting points at the vertices housing the rotator bearing and drive system. The rotator was driven by two servo motors via gear and pinion. The small diameter mounting point circle, the distance from the mounting points to major NFIRAOS frame members, and the cantilevered instrument load resulted in a low natural frequency, which led to concerns about earthquake survivability.

Just prior to the conceptual design the three mounting pads on NFIRAOS were moved out to a 6100 mm diameter circle to increase the support stiffness and stability. Keeping the rotator on the top of the instrument, and simply increasing the size of the support structure and moving the mounting points out, would have resulted in a large and heavy plate-like structure with poor stiffness. An initial support structure and rotator redesign (V2 in Figure 4) moved the rotator down to about the mid-point of the science dewar, supported by a large steel truss assembly. While this design addressed the stiffness, due to the large steel structure and large diameter rotator, it resulted in excessive mass.  An extensive weight reduction exercise then led to: the rotator being moved to a position between the OIWFS and Science Dewar (V3 in Figure 4) to enable its diameter to be reduced; a double truss support utilizing a carbon-fiber-reinforced polymer (CFRP) ring structure and tubular struts; and the cable wrap being detached from the instrument.

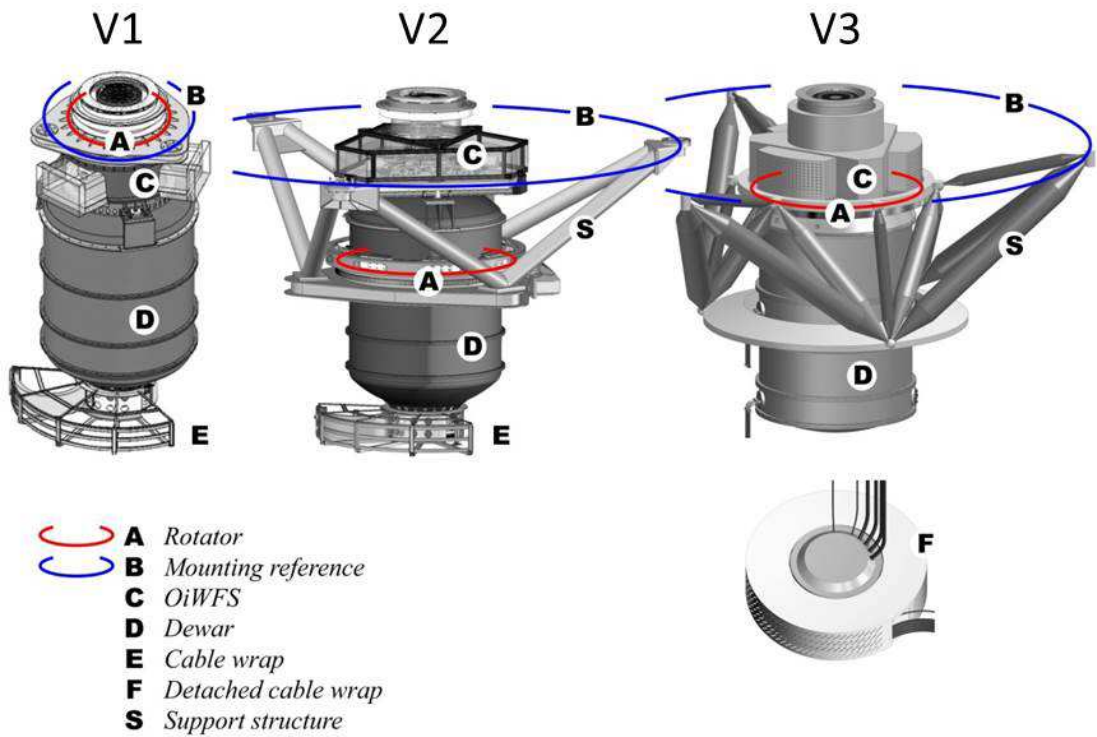

**Figure 4 - Evolution of the IRIS Support Structure/Rotator Design through three iterations (V1, V2 and V3).**

The support structure is a double hexapod concept with additional redundant struts for stiffness. The instrument load is transferred to the inner hexapod through the rotator stator. Two struts attach to each of three points on the stator forming the inner hexapod to transfer the load down to three points on a CFRP structural ring. From each of these points two larger diameter CFRP struts transfer the load up to each of the three NFIRAOS interface mounting points creating the outer hexapod. In order to increase the stiffness six more smaller-diameter CFRP struts are incorporated: three between the inner stator mounting points and the outer interface points, and three from the lower structural ring points to midpoints on the stator. Extensive use of CFRP is required to maintain stiffness while reducing mass.

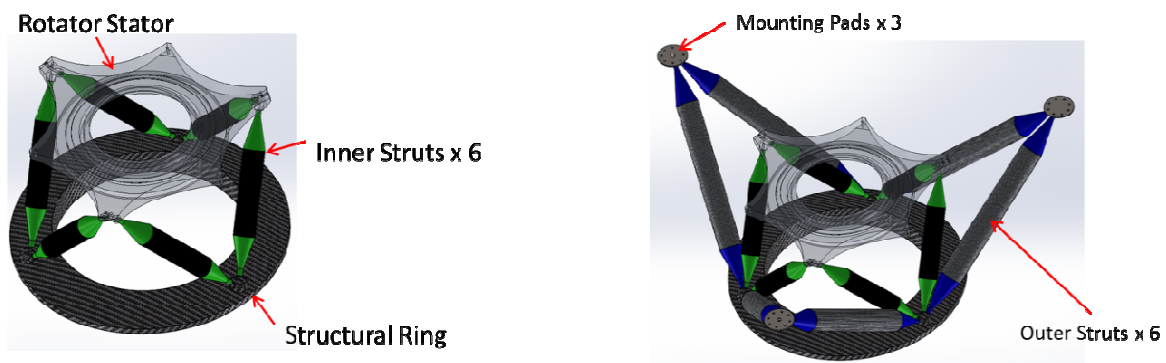

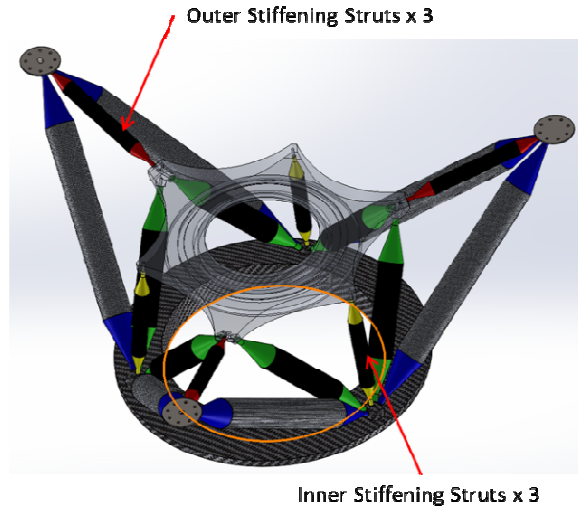

**Figure 5** – Figures illustrating different components of the re-designed support structure.

## 2.2 Rotator

The rotator is comprised of five components: the stator, the bearing, the hub, the motor, and an encoder. These components are further grouped into two subassemblies: stationary, and rotating. The stationary subassembly consists of the stator, outer bearing race, motor stator, and encoder read heads. The rotating subassembly consists of the hub, bearing inner race, motor rotor and encoder graduated strip.

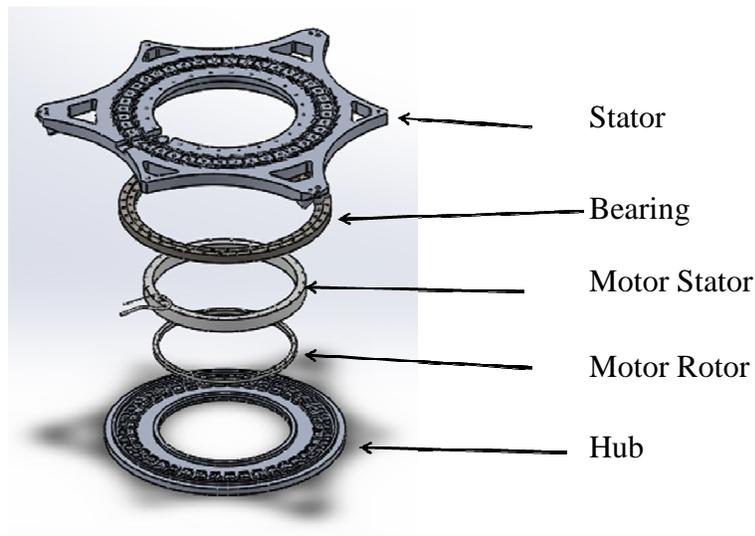

**Figure 6 - Rotator Assembly.**

The stator connects to the support structure struts, and six points around its circumference, and transfers the load to them from the outer race of the bearing. It is a large machined aluminum structure. Two concentric rings of flexures around the inner and outer race of the bearing mounting points provide compensation for differential expansion between the aluminum stator and steel bearing. The stator houses the motor stator and encoder read heads.

A Rollix crossed roller precision slewing bearing with a 1500mm pitch diameter has been chosen to provide low turning torque with high stiffness.

The rotator hub provides connections to the OIWFS and the Science Dewar, and transfers the loads to the bearing inner race. The motor rotor attaches to the hub to provide the drive torque for rotation. Flexures between mounting points provide compensation for differential expansion between the aluminum hub and steel bearing and motor rotor.

An Etel frameless, brushless torque motor was selected to provide smooth instrument rotation without the need for gearboxes or pinions and gears. The 1m diameter motor stator and rotator mount directly to the rotator stator and hub respectively. The motor has provisions for liquid cooling if it should be required to meet the stringent TMT waste heat allocations.

## 3. On-Instrument Wavefront Sensors, and On-Detector Guide Windows

AO corrections for IRIS are provided by NFIRAOS, which uses up to 6 laser guide stars, and one natural guide star, to detect (primarily high-order) atmospheric aberrations. Client instruments provide further low-order wavefront measurements, which in the case of IRIS, include TT and/or TTF measurements from the three OIWFS probes, and TT measurements from up to four ODGWs. Two deformable mirrors (DM) and a tip tilt stage within NFIRAOS are then used to correct the measured wavefront errors, providing a smooth (though curved) focal surface to IRIS. Light from NFIRAOS passes first through the OIWFS, and then enters the science dewar, where it ultimately illuminates the imager, and the IFS via a pick-off mirror. The rapid conversion of wavefront measurements to DM and tip tilt stage positional commands (as well as other offloading terms) is accomplished using a sophisticated Real-Time Controller (RTC) [3].

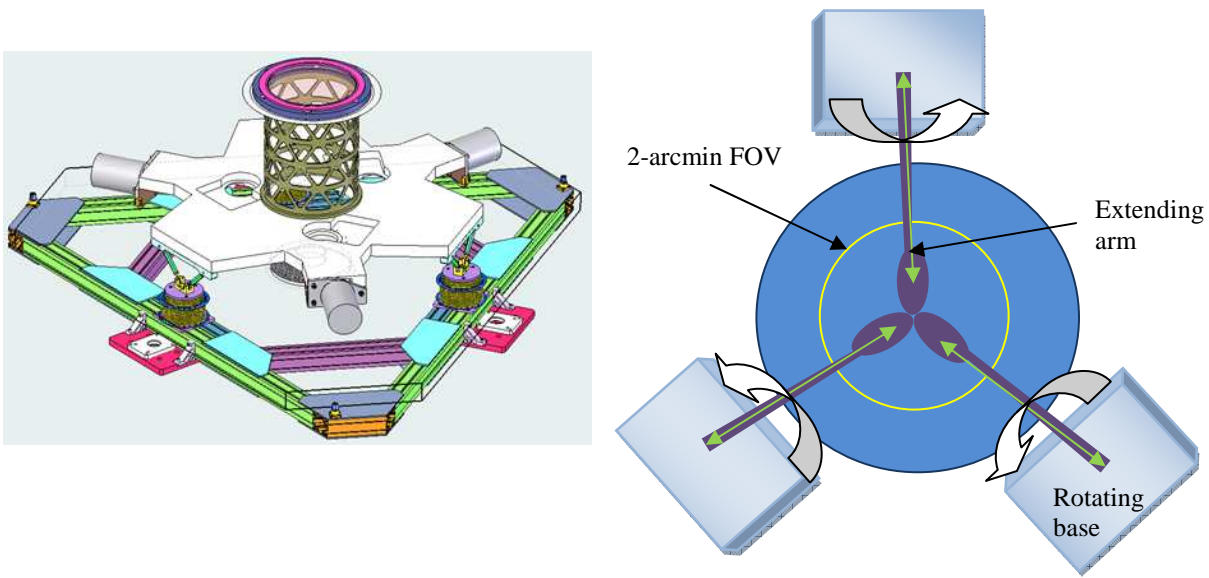

**Figure 7 –** *left:* **View of the OIWFS from the top without its enclosure.** *right:* **Illustration of OIWFS depicting the 2-arcmin FOV, and the degrees of freedom for the 3 probe arms.**

This section describes our proposed strategy for synchronizing the OIWFS and ODGW wavefront measurements with the RTC. The OIWFS probes, which patrol a roughly 2-arcmin field-of-view (FOV, see Figure 7), are configured for TT or TTF operations by switching between an imager lens, or a 2×2 Shack-Hartmann lenslet array. It is worth noting that, in order to obtain the best total sky coverage of viable guide star asterisms, the probes can operate within the same workspace (and may therefore collide); the collision avoidance algorithm is discussed in Ref. [1]. While the OIWFS detectors have a maximum 5-arcsec FOV, they can be configured to read-out smaller regions of interest (ROI) in order to accommodate faster readout times. These ROIs can also be reduced or enlarged, or tracked across the sensor dynamically while performing "on-chip guiding" (i.e., during real-time closed-loop AO operations of guide stars). OIWFS exposures are triggered by external pulses generated by NFIRAOS to ensure precise timing. Similarly, the IRIS imager detector may provide TT information by means of up to four ODGWs (one for each of its four imager detectors), which are restricted to configurable ROIs similar to the OIWFS. Complicating the ODGWs is the fact that ROI readouts, and NFIRAOS-generated triggers, must be configured and executed independently of the science exposures; these tasks are accomplished via specialized firmware that interleaves ODGW operations in the dead time between science row readouts. Finally, in order for the RTC to make use of OIWFS and ODGW pixel data, it must also send the:

- current angles of the OIWFS rotation stages and
- angle of the instrument rotator (in order to determine the orientation of the wavefront measurements with respect to the NFIRAOS DMs),
- currently-configured type of guiding (e.g., "on-chip" tracking of guide stars by moving the ROIs, or physically moving the probes in the case of the OIWFS).

### 3.1 Acquisition and Guiding

Acquisition refers to the initial detection of guide stars of the various wavefront sensors, and the process of reducing the sizes of their ROIs and/or readout rates, until the target guiding state is achieved. This acquisition process is performed by stepping through a pre-defined sequence of detector configurations that are kept in "acquisition and guiding tables". As both the OIWFS/ODGWs and NFIRAOS require knowledge of these steps, we plan to provide a common table to the RTC and IRIS. The tables themselves are generated dynamically (though asynchronously), as they depend on the current observing configuration (i.e., which types of wavefront sensors are to be used, the AO mode, and brightness of guide stars assigned to each probe). However, once received by the pertinent subsystems, the tables are stepped-through in real time, at the cadence of the NFIRAOS-generated triggers. This use of tables is currently the best working concept for coordinating the various actors given the operational requirements, and mechanisms for communication that are currently planned. This table will include information such as the frame rate, gains to apply for the step in question, and ROI window size. When the OIWFS/ODGW pixels are read, the most recently executed line of the table is included with the header sent to the RTC. As the RTC also has a copy of the same table, it will then have an accurate description of the pixels being received. Since all of these functions and data are used by the AO system, it is the AO Sequencer that coordinates these processes.

In the most typical case, the telescope will send a stream of targets to the OIWFS probe arms in order to follow guide stars. As described above, the RTC will also be sent a copy of the guide and acquisition table relevant to the current observation, such that it is able to accurately process and apply the correct pixel gains. This sequence of events is depicted in Figure 8.

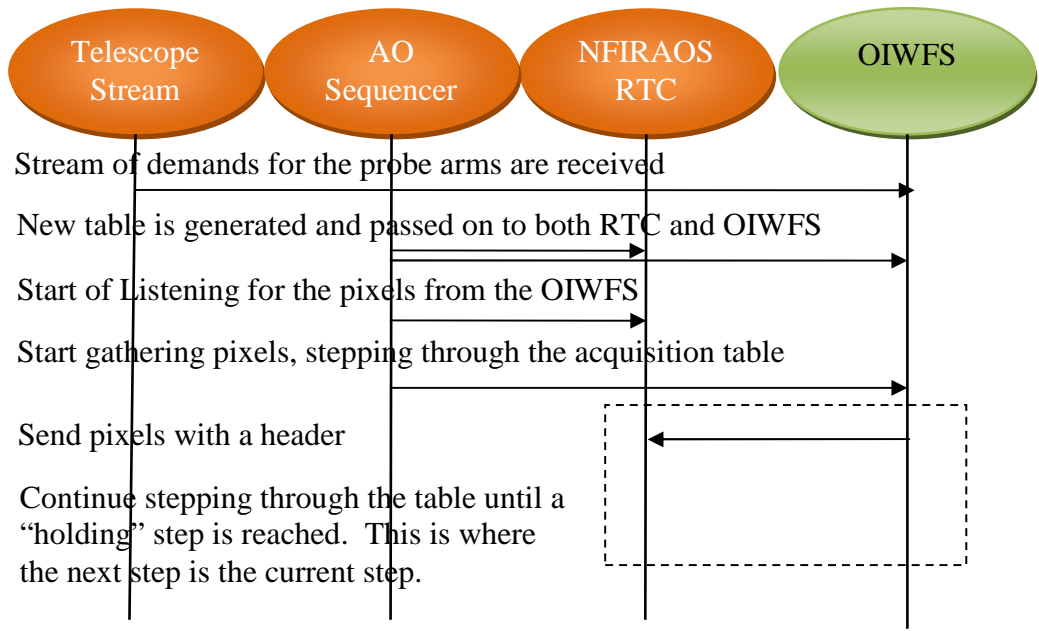

**Figure 8 - Acquisition Sequence for the OIWFS when following streams from the telescope.**

Another operational scenario we consider is the case of light being lost from a particular sensor during guiding (as seen by the RTC). In our current system architecture, the RTC reports this fact, and other subsystems (i.e., the AO Sequencer and OIWFS) are responsible for remedying the situation asynchronously (e.g., by moving the probes or ROI). However, during this time, in an effort to keep the AO system locked as long as possible, the guard pixels (i.e., a frame periodically exposed with a larger FOV) will be used to determine an appropriate step in the table to go to. If light can be detected in this way beyond the nominal ROI, it can decide to revert to an earlier step in the table with a larger window in real time. This appropriate step is passed on to the OIWFS, and the OIWFS can attempt to "reacquire" by once again stepping through the table from that point.

Further design efforts to enable the AO system to remain continually locked include handling telescope dithers. When dithering, the telescope moves small distances (a maximum of 30 arcsec), an event that can occur while wavefront corrections are being performed. This will move the guide stars across the focal plane substantially faster than during normal observations. Non-sidereal tracking can also be considered a type of slow dithering. In either of these situations, using the same table format, the windows can be enlarged, and/or the sample rates decreased.

Another similar case is that of on-chip guiding. This situation involves the tracking of guide stars across OIWFS detectors by moving the ROIs, but keeping the probe arms stationary. This functionality (and therefore the contents of the guide and acquisition tables) is very similar to that of the ODGWs for which only ROI positioning can be used to track stars.

### 3.2 ODGW Readout Considerations

As mentioned above, ODGW pixel readouts are generally interleaved with science exposures and readouts. This situation is further complicated when science exposures are not in progress; during these times the detectors are continuously being flushed (reset). As the RTC may require ODGW pixels during these periods, their readouts must continue despite the flush. Furthermore, the cadence of the ODGW readouts must remain constant (even when the RTC has not requested data), in order to minimize background changes induced in the chip by self-heating (a property of infrared imagers). These backgrounds become more problematic when the ROI is moved around, and handling position and time-dependent backgrounds will form part of the calibration strategy.

During acquisition, it may be the case that a larger guide window is required, and that there is insufficient time to read it in the dead time at the end of a single science row read. If that is the case, the guide window read will simply be split between the reads of two rows of science data. The details of these reads will be firmly established once the detector is available.

### 3.3 Alignment and Calibration

NFIRAOS, with input from the IRIS OIWFS, is required to deliver a wavefront error of 191nm within a 17"×17" FOV at the IRIS focal plane. Even though the OIWFS is designed to meet this requirement, we have been developing methods to mitigate potential sources of image blur and misalignment due to properties of the mechanical structure and interfaces. The sources of error include:

- NFIRAOS delivers a curved focal plane to IRIS. Boresight errors that arise when mounting IRIS on NIFRAOS will create focus errors across the field.
- The different coefficients of thermal expansion (CTE) of materials between NFIRAOS and the support structure can cause shifts in the IRIS structure with respect to NFIRAOS over long timescales.
- Imperfections in the rotator could induce nutation of the entire instrument which will in turn cause image blur.
- There may be movement due to temperature changes between the OIWFS and the Science Dewar, which would result in corresponding differential motion of the OIWFS and science focal planes.

We are developing strategies to calibrate and mitigate these potential sources of misalignment. NFIRAOS and the NFIRAOS Science Calibration Unit (NSCU) will contain multiple variable sources that can be used to characterize the performance and alignment of NFIRAOS to the OIWFS and the IRIS focal plane. We are also designing light sources that will mount on the underside of the OIWFS probes, which can be used to measure the relative positions and focus of the probe arms to the focal plane. The ODGW will also play a key role in mitigating the sources of error listed above.

## 4. Summary

Our goal during preliminary design is to produce a design that has retired or at least significantly mitigated all major risks while demonstrating that IRIS requirements are met and interfaces are well understood. In addition, we have taken on the significant task of reducing the mass of the support structure and rotator by 30 percent so that IRIS as a whole can meet its mass budget. To accomplish this goal while still being able to withstand seismic disturbances, a new design with composite materials was developed. To date, we have successfully reduced the weight of the rotator and structure by 27%. Our CSRO design will help enable near diffraction-limited performance and excellent sky coverage through three OIWFS that can be deployed in a large field of view and provide tip/tilt/focus and plate scale information to the NFIRAOS RTC.